\documentstyle[multicol,aps,prb,psfig]{revtex}

\def\beginwide{
        \end{multicols} \vspace*{-0.5cm} \noindent
        \rule{3.5in}{.1mm}\rule{.1mm}{5mm} \widetext \medskip }
\def\beginwidetop{
        \end{multicols} \vspace*{-0.5cm} \noindent
        \widetext \medskip }
\def\endwide{
        \hspace*{3.35in}~\rule[-5mm]{.1mm}{5mm}\rule{3.5in}{.1mm}
        \begin{multicols}{2} \vspace*{-1.0cm} \noindent }
\def\endwidebottom{
        \begin{multicols}{2} \vspace*{-1.0cm} \noindent }

\draft
\newcommand{\bsim}{\mbox{\raisebox{-0.1cm}{$\;
\stackrel{\textstyle>}{\sim}\;$}}}
\newcommand{\lsim}{\mbox{\raisebox{-0.1cm}{$\;
\stackrel{\textstyle<}{\sim}\;$}}}
\newcommand{\beq}{\begin{equation}}
\newcommand{\eeq}{\end{equation}}
\newcommand{\beqa}{\begin{eqnarray}}
\newcommand{\eeqa}{\end{eqnarray}}

\begin{document}

\title{Magnetic and lattice polaron in Holstein-$t$-$J$ model}

\author{E. Cappelluti$^1$ and S. Ciuchi$^2$}

\address{$^1$ Dipartimento di Fisica, Universit\`{a} di Roma
``La Sapienza'',
Piazzale A.  Moro, 2, 00185 Roma, Italy \\
and Istituto Nazionale Fisica della Materia, Unit\'a di Roma 1, Italy}

\address{$^2$ Dipartimento di Fisica, Universit\`{a} dell'Aquila,
v. Vetoio, 67010 Coppito-L'Aquila, Italy, \\
and Istituto Nazionale Fisica della Materia, Unit\'a dell'Aquila, Italy}

\date{\today}
\maketitle

\begin{abstract}

We investigate the interplay between the formation of lattice and magnetic
polaron in the case of a single hole in the antiferromagnetic background. We
present an exact analytical solution of the Holstein-$t$-$J$ model in infinite
dimensions. Ground state energy, electron-lattice correlation function, spin
bag dimension as well as spectral properties are calculated.  The magnetic and
hole-lattice correlations sustain each other, i.e. the presence of
antiferromagnetic correlations favors the formation of the lattice polaron at
lower value of the electron-phonon coupling while the polaronic effect
contributes to reduce the number of spin defects in the antiferromagnetic
background. The crossover towards a spin-lattice small polaron region of the
phase diagram becomes a discontinuous transition
in the adiabatic limit.
\\
PACS number(s): 71.10.Fd, 71.38.-k, 75.30.Kz, 71.38.Ht
\end{abstract}

\vskip 2pc


\section{Introduction}
\label{intro}

A single hole in an antiferromagnetic background represents a widely
studied problem in solid state physics due to its relevance to the problem
of high-$T_c$ superconductivity.\cite{dagotto,manousakis}
High-$T_c$ compounds indeed show an antiferromagnetic undoped phase
which is gradually degraded and finally destroyed by doping.
The electronic properties at low doping are therefore often described in terms
of Hubbard or $t$-$J$ model.\cite{anderson,zhang}

The problem of the motion of a single hole, far from being a pure academic
issue, can be representative of the extreme low doping case
of these materials.
It is worth to note that even in the case of a single hole the problem of
intermediate/strong magnetic interaction with an antiferromagnetic background
is a non trivial many-body problem. The difficulty consists in describing the
dressing of the hole by a cloud of spin background excitations
(the magnetic or spin polaron) which can
coherently moves as a quasiparticle.\cite{schmitt-rink,kane,eder}
The situation is similar to that of a small lattice polaron i.e. the case of
an electron moving together with a phonon cloud which represents the lattice
deformation induced by the presence of the charge.\cite{pines}

The connection between lattice and magnetic polaron goes however
beyond a methodological interest.
There are indeed several observations of a sizable interplay
between electron-phonon and magnetic interaction in
cuprates\cite{lanzara} as well as in
manganites.\cite{manga} Purpose of this paper is to explore in detail
the physical consequences of this interplay, in particular
with regards to the lattice and magnetic polaron properties.
To this aim a non perturbative way is clearly needed.

We present an exact analytical solution of the Holstein-$t$-$J$ model for a
single hole in infinite dimension.
Ground state energy, electron-lattice
correlation function, spin bag dimension as well as spectral properties are
calculated. We find that the lattice (spin) polaron formation
depends strongly on the magnetic (hole-phonon) interaction.
We identify thus regions of phonon assisted magnetic
 polaron as well
as magnetic induced lattice polaron. The extension of these regions
are strongly dependent on the adiabatic ratio and they vanish in the adiabatic
limit. In that regime lattice and magnetic polaron formations are strictly
tied each other.
These general results could help to explain the
strong interplay between lattice and spin degrees of freedom in cuprates
and in manganites. Finally we
discuss the major drawback of our approximation and we give some ideas to
overcome it.

\section{The Holstein-\lowercase{$t$}-$J$ model and its DMFT solution}
\label{model}

Let us consider the Holstein-$t$-$J$ model defined by the
Hamiltonian:\cite{martinez,ramsak}
\begin{eqnarray}
H&=&-\frac{t}{2}\sum_{\langle ij \rangle \sigma}
\left(\tilde{c}^\dagger_{i \sigma}\tilde{c}_{j \sigma} + {\rm h.c.}\right)
\nonumber\\
&&+\frac{J}{2}\sum_{\langle ij \rangle}
\left[\left(S_i^z S_j^z -\frac{n_in_j}{4}\right)
+\frac{1}{2}\left(S_i^+ S_j^- + S_i^- S_j^+ \right)\right]\nonumber\\
&&+g \sum_i  n_i  (b_i+b_i^\dagger)
+\omega_0\sum_i b_i^\dagger b_i,
\label{hamil1}
\end{eqnarray}
where $\tilde{c}^\dagger_{i \sigma}$ are the electron operators
in the presence of infinite on-site repulsion that prevents
double occupancy [$\tilde{c}^\dagger_{i \sigma}
= c^\dagger_{i \sigma}(1-n_{-\sigma})$],
$b^\dagger_i$ the phonon operators and
$S^{z,+,-}_i$ respectively the $z$ component
and the raising and lowering spin operators.
The first term in Eq. (\ref{hamil1}) describes the hopping of the electrons
on nearest neighbors of a square lattice,
the second one the direct and the exchange interaction,
the third one the local electron-phonon interaction
coupled to charge density, and the fourth the Einstein
phonon frequency.
The choice of a hopping matrix element equal to $t/2$ gives rise
to a band with bare bandwidth $t$.
The model can be straightforward generalized in infinite dimensions
by using the usual rescaling:
$t \rightarrow t/\sqrt{z}$, $J \rightarrow J/z$, where
where $z$ is the coordination number.
For a hypercubic lattice the coordination number is $z=2d$,
while for a Bethe lattice $z=d$.

All through out this paper we shall consider one hole created on the
anti-ferromagnetic half filled state.
Antiferromagnetic state is described in terms of a {\it classical}
N\'eel ground state.
A convenient approach to this aim is the spin wave
theory applied to lattice model as can be found e.g.
in Ref.~\onlinecite{schmitt-rink,kane,martinez}.
A useful effective Hamiltonian can be thus derives
by mainly following the discussion in Ref.~\onlinecite{martinez},
generalized now in the presence of a Holstein electron-phonon interaction.
The Hamiltonian is first transformed by
a canonical transformation into
a ferromagnetic one.
Then ``hole'' and ``spin'' defect operators
are introduced, respectively
as fermionic $h$ and bosonic $a$ operators on the antiferromagnetic
ground state. The resulting Hamiltonian reads thus:
\begin{eqnarray}
H&=&\frac{t}{2\sqrt{z}}\sum_{\langle ij \rangle}
\left(h_j^\dagger h_i a_j + {\rm h.c.}\right)
\nonumber\\
&&
-g \sum_i h_i^\dagger h_i
(b_i+b_i^\dagger)
+\omega_0\sum_i b_i^\dagger b_i\nonumber\\
&&+\frac{J}{4z}\sum_{\langle ij \rangle}
\left[a_i^\dagger a_i+a_j^\dagger a_j
+a_i^\dagger a_j^\dagger+a_i a_j\right]\nonumber\\
&&-\frac{J}{2z} \sum_{\langle ij \rangle} h_i^\dagger h_i a_j^\dagger a_j
-\frac{J}{2z} \sum_{\langle ij \rangle} a_i^\dagger a_i a_j^\dagger a_j
\nonumber\\
&&+\frac{J}{2} \sum_i h_i^\dagger h_i-\frac{J}{4},
\label{hamilha}
\end{eqnarray}
where we have neglected the hole-hole terms
since we are interested in a single hole in an antiferromagnetic background.

Eq.~(\ref{hamilha}) can be significantly simplified in infinite
dimension. In that limit indeed the two terms of the fourth line can be
shown to be negligible since they contribute only at $O(1/d)$.
In addition, in the absence of any boson condensate
$\langle a \rangle$, $\langle a^\dagger \rangle$, which should destroy
the antiferromagnetic background and which is forbidden
in our context, also the last two terms of the third line
can be dropped. We end up thus with
the effective Hamiltonian of the Holstein-$t$-$J$ model valid in infinite
dimension:
\begin{eqnarray}
H&=&\frac{t}{2\sqrt{z}}\sum_{\langle ij \rangle}
\left(h_j^\dagger h_i a_j + {\rm h.c.}\right)\nonumber\\
&&-g \sum_i h_i^\dagger h_i
(b_i+b_i^\dagger)
+\omega_0\sum_i b_i^\dagger b_i\nonumber\\
&&+\frac{J}{4z}\sum_{\langle ij \rangle}
\left[a_i^\dagger a_i+a_j^\dagger a_j\right]
+\frac{J}{2} \sum_i h_i^\dagger h_i.
\label{hamilhatjz}
\end{eqnarray}
The first term of Eq. (\ref{hamilhatjz}) describe the kinetic
hopping of one hole
on the antiferromagnetic background, which is accompanied by the creation
(destruction) of a spin defect which breaks (restores)
$2z$ magnetic bonds with individual energy $J/4z$.
In addition
we have the usual local electron-phonon interaction which couples
{\rm the hole density} to the local phonon.
The last term in Eq. (\ref{hamilhatjz}) can be absorbed in the
definition of the hole chemical potential which, for the single
hole case here considered, has to be set at the bottom of the hole band.
It is important to note that, although
written in terms of hole operators,
the phonon part (free phonon part +
hole-phonon interaction) is formally identical to the Holstein model.
This is not a trivial result since in a half-filling case
all the electrons are coupled with the phonons and one
should in principle deal with a many-body problem.
As a consequence we are thus able
to reduce the many-body problem to a single-particle (one hole) system
interacting with phonons and with spin defects through
Eq. (\ref{hamilhatjz}).

Although the Hamiltonian (\ref{hamilhatjz})
looks much more affordable than (\ref{hamil1}), the analytic study
of its properties is still a quite hard task at finite dimension.
The problem can be much simplified however in infinite dimensions
where, as we are going to see,
the limit of infinite coordination number $z \rightarrow \infty$,
\cite{muller-hartmann,metzner,georges,ciuchi}
all together with the retraceable path constraint enforced by the
antiferromagnetic background,\cite{strack} provides an {\em exact} solution.
A explicit derivation can be found in Appendix \ref{appfrac}.
In this section we only summarize the final equations which determine
in a self-consistent way the hole Green's function.

A crucial point is the possibility of writing
the self-energy of the
local propagator
as the sum of two contributions, labelled
as $\Sigma_{\rm hop}(\omega)$ and $\Sigma_{\rm el-ph}(\omega)$,
which closely resemble the functional expressions of the
hopping and phonon self-energy respectively in the pure $t$-$J$ and
Holstein models, but {\em which are now evaluated in the presence
of both exchange and phonon interactions}.
We can write thus:
\begin{equation}
\label{eqG}
G(\omega) = \frac{1}{\omega - \Sigma_{\rm hop}(\omega)-
\Sigma_{\rm el-ph}(\omega)},
\end{equation}
where the ``hopping'' contribution is given by\cite{strack}
\begin{equation}
\label{Sigmat}
\Sigma_{\rm hop}(\omega)  = \frac{t^2}{4} G(\omega-J/2),
\label{sigmat}
\end{equation}
and the ``phonon'' self-energy can be expressed
by means of a continued fraction:\cite{ciuchi}
\begin{equation}
\label{Sigmaep}
\Sigma_{\rm el-ph}(\omega)=\frac{g^2}{
G^{-1}_t(\omega-\omega_0)-\frac{\displaystyle 2g^2}{
\displaystyle G^{-1}_t(\omega-2\omega_0) -
\frac{\displaystyle 3g^2}{
\displaystyle G^{-1}_t(\omega-3\omega_0) -
\ldots
}
}
},
\end{equation}
where
\begin{equation}
\label{Gt}
G^{-1}_t(\omega)=\omega-\frac{t^2}{4} G(\omega-J/2).
\label{gt}
\end{equation}

It should be stressed again that both $\Sigma_{\rm hop}(\omega)$ and
$\Sigma_{\rm el-ph}(\omega)$ are functions of the {\em total}
Green's function $G$ [Eqs. (\ref{sigmat})-(\ref{gt})] which
contains the full dynamics (hopping, exchange, electron-phonon)
of the system. This not trivial self-consistency
accounts thus for the complex interplay between the phonon
and spin degrees of freedom.

Eqs. (\ref{eqG})-(\ref{gt}) represent a closed self-consistent system
which we can be numerically solved by iterations to obtain the explicit
{\em exact} expression of the local
Green's function $G(\omega)$, and hence
any local one-particle relevant property of the system.

The formal scheme looks quite similar to the dynamical
mean field theory in infinite dimension for a Bethe lattice, applied
for instance at the purely electron-phonon system.\cite{ciuchi}
However, due to the antiferromagnetic background,
the physical interpretation is quite different.

Due to the orthogonality of the initial and final
antiferromagnetic background, the non-local component
of the Green's function in the Holstein-$t$-$J$ model for $J=0$
is strictly zero $G_{ij}(\omega)=G(\omega)\delta_{i,j}$,\cite{strack} whereas
for the pure Holstein model $G_{i \neq j}(\omega)$ is finite and
provides information about the non-local
dynamics:
$G({\bf k},\omega)=1/[\omega -\epsilon_{\bf k}-\Sigma(\omega)]$.

In addition, the magnetic ordering has important consequences
also on the local Green's function $G_{ii}(\omega)$.
In the pure Holstein model for instance $G_{ii}(\omega)$ takes
into account any generic dynamics which occurs back and forth a given site
whereas
in the Holstein-$t$-$J$ model the electron must follow
a retraceable path in order to restore the antiferromagnetic
background.\cite{strack}
A Bethe-like dynamics is thus enforced by the magnetic ordering regardless
the actual real space lattice.
The object made up by the
hole plus the local modification of the spin configuration due
to the presence of the hole is the ``spin polaron''.

The local constraint $G_{ij}(\omega)=G(\omega)\delta_{i,j}$ induced
at $d=\infty$
in the Holstein-$t$-$J$ model by the antiferromagnetic background
can appear a quite strong simplification.
However, it should be reminded that it holds true as long as
the antiferromagnetic spin configuration
can be assumed to be frozen, in particular,
as long as the Hamiltonian does not induce spin dynamics.
This is the case of the effective Holstein-$t$-$J$ Hamiltonian
(\ref{hamilhatjz}) in infinite dimension where spin fluctuations
are neglected.\cite{strack}
The existence of the spin polaron itself could be questioned
at finite dimension where spin fluctuations are operative.
However, several numerical and analytic studies have shown
that the restoring of spin fluctuations does not destroy the spin polaron
object, but opens coherent channels of hole
propagation.\cite{schmitt-rink,kane,eder}
In this situation the spin polaron can thus propagate as a whole through the
crystal.

Many studies have investigated the motion of the spin polaron
and determined its ${\bf k}$-dispersion and optical conductivity,
both in the absence\cite{martinez,dagotto2,poilblanc,leung}
and in the presence of electron-phonon
interaction.\cite{ramsak,wellein,kyung,bauml,moskalenko,capone}
However, apart few
exceptions,\cite{barnes,wellein} the internal degrees of freedom of
the spin (lattice) polaron object have not so far been much investigated.
In the present work we mainly focus on the formation of
the spin polaron and on its internal structure properties (size,
binding energy etc\ldots). These quantities are local features
which, we believe, are only weakly affected by the itinerant nature of the spin polaron.
In this perspective we think that the infinite dimension approach
here considered provides valuable information on the spin polaron
formation and on its interplay with the local (Holstein)
electron-phonon interaction.

\section{Spectral properties}
\label{spectral}
In order to have a complete description of the physical properties
of the Holstein-$t$-$J$ model, it is useful to identify three
independent dimensionless parameters:\cite{feinberg}
the adiabatic ratio $\omega_0/t$,
the electron-phonon coupling $\lambda =  g^2/\omega_0 t$ and the exchange
interaction $J/t$. Another important parameter to be defined
is the multiphonon constant $\alpha=g/\omega_0$ that is a local quantity
which does not involves electron hopping $t$.
Limiting case are $\lambda = 0$ where the system reduces to the
$t$-$J$ model, and $J/t = 0$ where the Holstein model
{\it on an antiferromagnetic background} is recovered.

A naive look could regard this problem as a simple interplay
between two energy scales: the exchange $J/t$ which rules the magnetic
properties, and $\lambda$ related to the electron-phonon coupling.
The dominance of one of them would therefore determines the overall
properties of the system, while the weaker one could be considered
as a perturbation. However, as we are going to see,
this picture is too simplistic.
A more accurate description of the physics must take into account first of all
the role of the adiabatic ratio ($\omega_0/t$) which in the pure Holstein model
rules also the polaron crossover.\cite{capone2}
In particular, we can expect that the magnetic (lattice)
polaron formation induces a drastic renormalization  of the kinetic
energy scale $t$. The ``effective'' adiabatic ratio will depend in an implicit
way on the electron-phonon and magnetic interactions.
Then the magnetic (lattice)
polaron formation induces a drastic renormalization  of the kinetic
energy scale which in its turn affects the phonon (magnetic) properties
and eventually leads toward intermediate/strong hole-phonon couplings.
It is interesting to note that the multiphonon parameter $\alpha=g/\omega_0$
does not depend on $t$, so that it can be considered unaffected
by local hopping renormalization (we remind that phonon frequency
$\omega_0$ is not screened by electron-phonon interaction for the single
hole case).

Let us now first discuss spectral properties
by studying the spectral density defined
as $A(\omega)=-(1/\pi) \mbox{Im}[G(\omega)]$ directly accessible by the
knowledge of Green
function.

All through this paper we consider a Bethe lattice a the standard
semicircular density of states with bandwidth $2t$. It should be noted however
that, since the antiferromagnetic background enforces a retraceable
path approximation, a semicircular density of states is recovered
independently of the chosen crystal lattice, i.e.
also for a hypercubic lattice. In other words Eqs. \ref{eqG}-\ref{Gt} are valid
for any lattice structure provided the coordination number is infinite.
The assumption of a non-retraceable
Bethe lattice however allows to classify explicitly all the
hopping processes which leads to the magnetic polaron
formations.\cite{strack} Moreover in the Bethe lattice, contrary to
hypercubic infinite bandwidth
case, the single polaron problem can be well defined.\cite{ciuchi}
To obtain the spectral function we have iterated Eqs. (\ref{eqG})-(\ref{Gt})
up to numerical convergence by using a truncation in the continued
fraction Eq. (\ref{Sigmaep}) according to the procedure outlined in
Ref.~\onlinecite{ciuchi} i.e. we truncate the continued fraction Eq.
(\ref{Sigmaep}) at a stage $N_{\rm ph} \gg \alpha^2$.
Moreover we notice that a continued fraction arises also in the
absence of hole-phonon interaction\cite{strack} once $G_t$
(Eq. (\ref{Gt})) is substituted in Eq. (\ref{eqG}).
We have found that in order to properly generate a certain number $M$
of the magnetic poles of the spectrum we have to choose a truncation
$N_J \gg M$.

\begin{figure}
\centerline{\psfig{figure=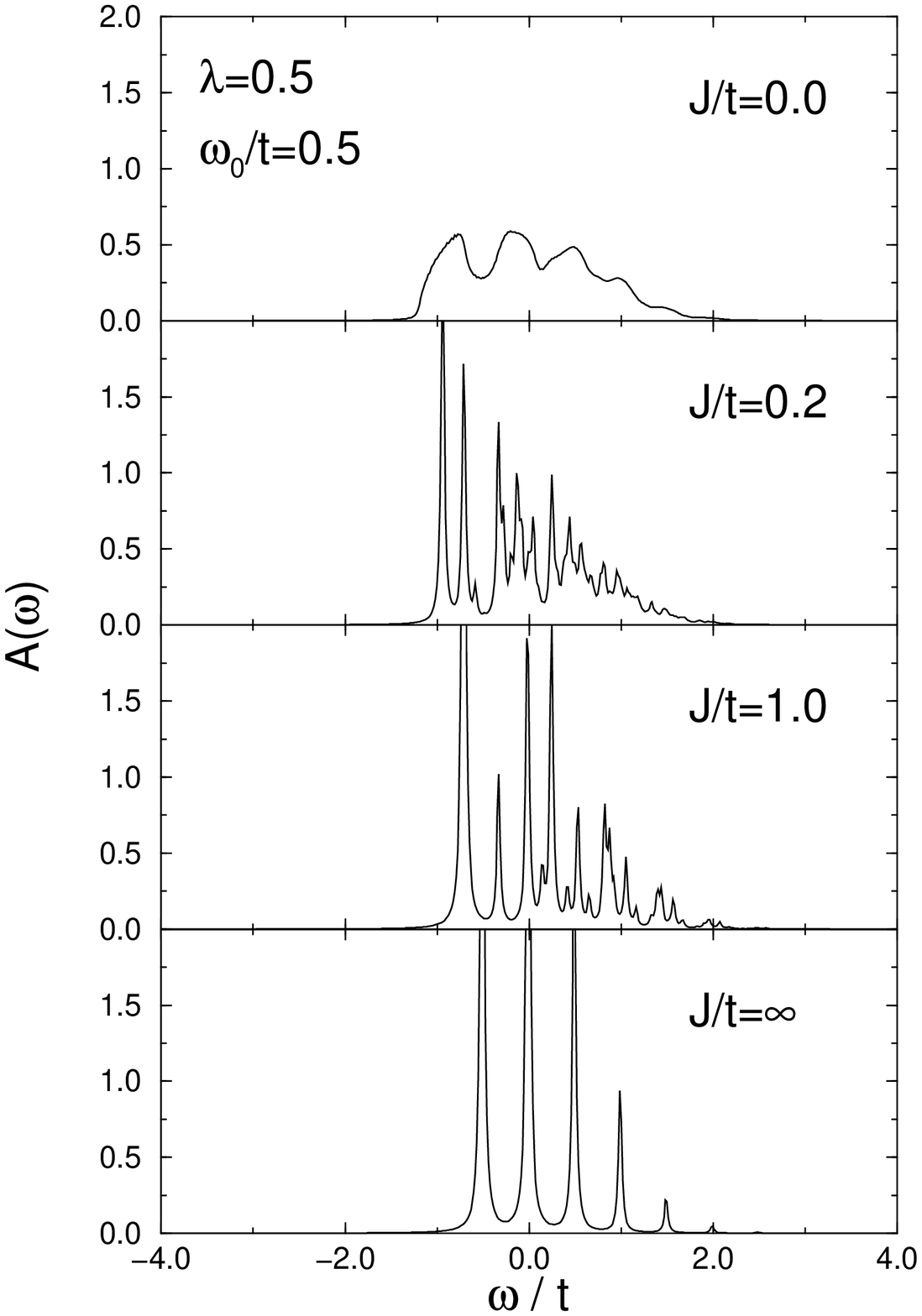,width=7cm,clip=}}
\narrowtext
\caption{Spectral density $A(\omega)$ for different values of
exchange energy: $J/t = 0, 0.2, 1.0, \infty$ and $\lambda=0.5$,
$\omega_0/t=0.5$.
We use a finite broadening ($0.02 t$) is used.}
\label{fsp-g=05-l=05}
\end{figure}

In Fig.~\ref{fsp-g=05-l=05} we show the evolution of the spectral
density for moderate values of $J/t$ and $\lambda=0.5$, $\omega_0/t=0.5$.
For $J/t=0$ the spectral density is made
by a continuum with incipient structures due to the electron-phonon coupling.
This shape is characteristic of weak electron-phonon coupling
in a small-intermediate adiabatic regime $\omega_0/t \lsim 1$,
with no well defined polaron peak.\cite{ciuchi}
By switching on the exchange interaction $J/t=0.2$ the continuum spectrum
is split in a set of magnetic peaks.
This trend is quite similar to what happens in the infinite
dimensional $t$-$J$ model\cite{strack}
with an additional modulation due to the underlying
electron-phonon features.
We can thus think of the resulting spectral function as ruled by different
couplings on different scale, where the gross structure on scale $t$
is determined by the pure electron-phonon interaction superimposed
by the fine structure on scale $J$ given by the magnetic peaks
spaced as $(J/t)^{2/3}$ (for small $J/t$).
By further increasing $J/t$ we pass an intermediate regime where
magnetic and phonon peaks are mixed together and eventually
we have a purely phononic spectrum with a set of peaks
starting from $-g^2/\omega_0$ and
equally spaced by $\omega_0$.
This is indeed
just the characteristic spectrum of the atomic Holstein model
with $\alpha^2=0.5$.\cite{mahan}
As discussed above, the origin of such a behavior
is the strong hopping amplitude renormalization due
to the magnetic polaron formation for $J > 1$, which drives the
system from an almost adiabatic case to an effective
antiadiabatic regime $\omega_0/t^* \gg 1$.
It is however surprising that a strong magnetic interaction
$J/t \gg \lambda$ yielded a purely phononic spectrum!.

\begin{figure}
\centerline{\psfig{figure=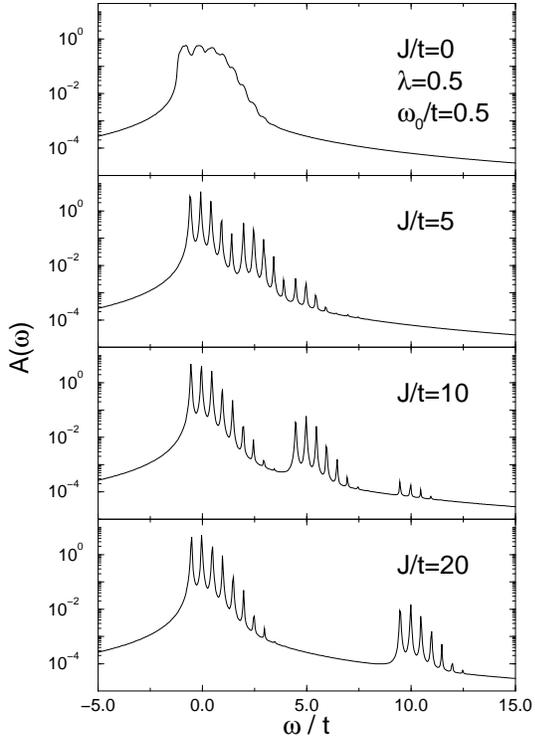,width=7cm,clip=}}
\narrowtext
\caption{Spectral density
$A(\omega)$ for different values of
exchange energy: $J/t = 0, 5, 10, 20$ and $\lambda=0.5$,
$\omega_0/t=0.5$.}
\label{fsp-g=05-l=05r}
\end{figure}

In order to better understand the global
evolution of the spectral function it is interesting to look at
what happens on a larger energy scale.
In Fig. \ref{fsp-g=05-l=05r} we have thus plotted the
the behavior of $A(\omega)$ on a linear-log scale for
$J/t=0,5,10,20$.
We see that the spectrum for finite $J/t$ is roughly determined
by replicas of the $J/t=0$ spectrum equally spaced by $J$ and
with vanishing spectral weight.
On a closer look we find that the structure can be derived
from the spectrum of the
pure $t$-$J$ model, made for $J/t \gg 1$ by
equally spaced magnetic peaks $\sum_n a_n \delta(\omega - n J/2)$,
by broadening each peak according the electron-phonon interaction
as in $J/t=0$ case.
The inverse occurs in
for $J/t < \lambda$. In this case we found magnetic structures
on a small energy scale $J$
coexisting with a phonon structure spread on energy $t$.
On the contrary the spectral function for large $J/t$
can be thought as built by pure magnetic structure for large energy
superimposed to a finer phononic structure made by peaks separated by the bare
phonon frequency $\omega_0$. This feature resembled the ``interband'' transitions
found in Ref.~\onlinecite{bauml} in the anti-adiabatic regime.
Of course when the magnetic coupling strictly goes to infinite
$J/t \rightarrow \infty$
the high energy magnetic peaks are shifted to infinite energy and
completely loose their spectral weight. The spectral function reduces
in this case to what shown
in the lower panel of Fig.~\ref{fsp-g=05-l=05}.

\section{Ground state properties}
\label{Ground state}

Ground state properties can be derived in a direct way by the knowledge
of the Green's function which allow the evaluation of the ground state
energy $E_0$ through the determination of the lowest band edge or the
lowest pole.

The character of the ground state can be determined by the knowledge of
several relevant quantities that can be evaluated using the
Hellman-Feynman theorem as:\cite{feynman}
\begin{itemize}
\item[i)] The mean number of phonons
$N_{\rm ph}=\langle b^\dagger b \rangle$:
$N_{\rm ph} = \partial E_0 / \partial \omega_0$;
\item[ii)] The mean hole-phonon correlation function
$C_0=\langle h^\dagger h (b+b^\dagger)\rangle$:
$C_0=-\partial E_0 / \partial g$;
\item[iii)] The mean number of spin defects
$N_{\rm s.d.}=\langle a^\dagger a \rangle$:
$N_{\rm s.d.}= 2 \partial E_0 / \partial J$;
\item[iv)] The effective
hopping amplitude $t^*$:
$t^*/t = 2 \partial |E_0| / \partial t$.\cite{note2}
\end{itemize}

Quantities i) and ii) shed light on the lattice polaron formation
process. A sharp increase of $C_0$ ($N_{\rm ph}$) is expected around some
intermediate value of the hole-phonon coupling $\lambda$
in the adiabatic regime.
$C_0$ increase from zero to its strong coupling limit
$2 \alpha$ ($\alpha^2$). The transition becomes a crossover
which becomes smoother and smoother approaching
the antiadiabatic case.\cite{ciuchi,capone2}

Quantity iii) provides information on the size of the magnetic
polaron.\cite{barnes}
In fact, since the retraceable path approximation is enforced by
the antiferromagnetic background in infinite dimension, it is clear
that the $N_{\rm s.d.}$ gives the {\em length} of the string of
spin defects\cite{shraiman,eder,strack}
which, in the Bethe lattice, is also the {\em size}
of the magnetic polaron. In the zero exchange limit
$J \rightarrow 0$ no energy cost is associated with a spin defect and
the size of the magnetic polaron diverges (large magnetic polaron limit).
In the $J \rightarrow \infty$ case instead spin defects are unfavored
and the magnetic polaron becomes pretty local (small magnetic polaron limit).
It could be appear surprising that information of non local quantities,
as the magnetic polaron size, could be available in the local
approach we are using. However it should be reminded that this is only
a mean quantity. The price we are paying by using
the local approximation exact in infinite dimensions is the impossibility
to have information on the probability distribution to find a path
with a given length $n$. We shall see later that a careful inspection
of the dependence of $N_{\rm s.d.}$ on the magnetic coupling $J$
can nevertheless provide valuable information about the
rough shape of the magnetic polaron size distribution.

In order to establish however a criterion for the
large-small magnetic polaron formation, we define the value
$N_{\rm s.d.}=0.5$ as the conventional transition between large and
small magnetic polaron. We can indeed think that for $N_{\rm s.d.}>0.5$
the probability to find paths with length $n \geq 1$ is larger
that the probability to have a completely magnetically trapped
hole ($n=0$), and we are therefore dealing with ``large''
magnetic polarons.

Finally the tendency to localization that can be due either
to magnetic or hole-phonon interaction can be deduced from the behavior of
the effective hopping iv).

Returning to the analysis of the lattice polaron case
we could certainly use as well the criterion
$N_{\rm ph} \geq 1$ or $C_0 \geq 1$ to identify the
lattice polaron
formation.\cite{ciuchi,wellein,capone,feinberg,capone2}
However, because of the local nature
of the electron-phonon interaction, we can gain a deeper insight
by looking at the {\em probability distribution} of
the phonon numbers.\cite{wellein,bauml}
The probability distribution of phonon numbers is defined as
\begin{equation}
\label{defPn}
P(n)=|\left< n|h|0\right>|^2
\end{equation}
where $|n\rangle$ is a state with zero holes and $n$ phonons and
$|0\rangle$ is the ground
state of the single hole. It can be obtained as a residue at the ground
state energy $\omega=E_0$ of $G^{nn}(\omega)$ [Eq. (\ref{green-nm})].
From the same  procedure outlined in Appendix \ref{appfrac}
(see also Ref.~\onlinecite{ciuchi})
one obtains:\cite{corrige}
\begin{equation}
\label{Gnn}
G^{nn}(\omega)={1 \over\displaystyle \omega-n\omega_0-\Sigma_{\rm hop}(\omega-n\omega_0)
-\Sigma_{\rm em}(\omega)-\Sigma_{\rm abs}(\omega)},
\end{equation}
where $\Sigma_{\rm em}(\omega)$ represents the processes related to the
emission of a phonon from the state $|n\rangle$,
\begin{equation}
\label{Sigma_em}
\Sigma_{\rm em}(\omega)=
{\strut (n+1)g^2 \over\displaystyle G^{-1}_t(\omega-n\omega_0-\omega_0)-
{\strut (n+2)g^2 \over\displaystyle G^{-1}_t(\omega-n\omega_0-2\omega_0)-
{\strut (n+3)g^2 \over\displaystyle G^{-1}_t(\omega-n\omega_0-3\omega_0)-
\cdots}}},
\end{equation}
and $\Sigma_{\rm abs}(\omega)$ takes into account the absorption processes which
are allowed also at zero temperature by the initial $n$-phonon state
\begin{equation}
\label{Sigma_abs}
\Sigma_{\rm abs}(\omega)=
{\strut ng^2 \over\displaystyle G^{-1}_t(\omega-n\omega_0+\omega_0)-
{\strut (n-1)g^2 \over\displaystyle G^{-1}_t(\omega-n\omega_0+2\omega_0)-
{\strut (n-2)g^2 \over\displaystyle \ddots -
{\strut g^2 \over\displaystyle G^{-1}_t(\omega)}}}}.
\end{equation}
The propagator $G^{-1}_t(\omega)$ is defined
in Eq. (\ref{Gt}), so that $G^{nn}(\omega)$ is a direct by-product
of the self-consistent solution of Eqs. (\ref{eqG})-(\ref{Gt}).
From that we immediately obtain the phonon
number distribution as
\begin{equation}
\label{eqPn}
P(n)=\left( \frac{\partial \left[G^{nn}(\omega)\right]^{-1}}
{\partial \omega}
.
\right)^{-1}_{\omega=E_0}.
\end{equation}

For a noninteracting system ($g=0$) the phonon distribution
contains only a $\delta$-peak at $n=0$: $P(n)=\delta_{n,0}$.
Switching on the electron-phonon interaction, the onset of
local lattice distortions are reflected in a shift of total weight towards
higher multiphonon peaks.
We can now unambiguously identify the lattice polaron formation
with the condition
\begin{equation}
P(n=0) \leq P(n=1).
\label{lpcrit}
\end{equation}
It is easy to check that this definition reproduces the well known
results for the Holstein model, namely the criterion
$\lambda \geq \lambda_c$ in the adiabatic limit
($\lambda_c = 0.844$), and $\alpha \geq 1$ in the antiadiabatic
one.\cite{feinberg,capone2,ciuchi} It could be worth to remark that the polaron
transition occurs however in a different way in the two limits.
In the nonadiabatic regime the most probable phonon number
$\bar{n}$ evolves in a smooth way by increasing $\lambda$
from $\bar{n}=0$ to higher
numbers $\bar{n} \sim \alpha^2$.
The dependence on $\lambda$ becomes sharper and sharper
by decreasing the adiabatic ratio $\omega_0/t$ and in the
adiabatic limit $\omega_0/t=0$ $\bar{n}$ jumps in a discontinuous way
from $\bar{n}=0$ for $\lambda < \lambda_c$
to $\bar{n}=\infty$ for for $\lambda > \lambda_c$.

\begin{figure}
\centerline{\psfig{figure=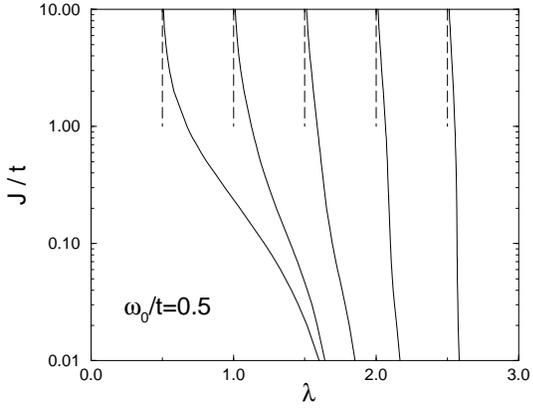,width=7cm,clip=}}
\narrowtext
\caption{Multiphonon processes in the $\lambda$-$J/t$ space
for $\omega_0/t=0.5$. From the left to the right the lines
correspond to $P(n=1) \geq P(n=0)$, 
$P(n=2) \geq P(n=1)$, $P(n=3) \geq P(n=2)$, etc.. 
Dashed lines indicate the antiadiabatic limit, respectively
$\alpha^2 = 1, 2, 3$, etc..}
\label{multiphonons}
\end{figure}

In Fig.~\ref{multiphonons} we plot the transition curves
corresponding respectively to $P(n=1) \geq P(n=0)$,
$P(n=2) \geq P(n=1)$, $P(n=3) \geq P(n=2)$, etc. in the $\lambda$-$J/t$
phase diagram for $\omega_0/t=0.5$.
Lattice polaron formation occurs on the left line corresponding
to $P(n=1) = P(n=0)$. We notice a strong dependence of the
lattice polaron formation on the magnetic energy.
In particular increasing the exchange coupling $J/t$ the
lattice polaron formation is shifted to smaller values of
$\lambda$ until at $J/t \rightarrow \infty$ lattice polaron
formation is ruled by the antiadiabatic criterion $\alpha^2 =1$.

\begin{figure}
\centerline{\psfig{figure=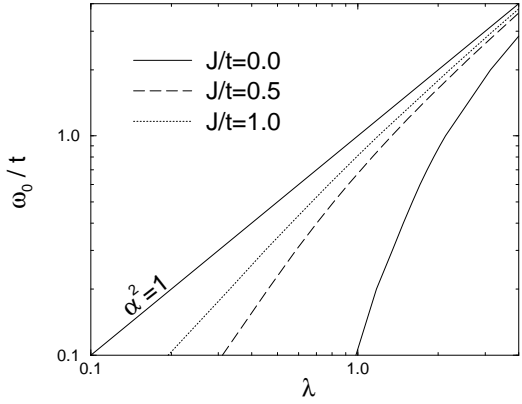,width=7cm,clip=}}
\narrowtext
\caption{Lattice polaron transition as determined by Eq.~(\ref{lpcrit})
in the $\lambda$-$\omega_0/t$ space for different exchange couplings:
$J/t=0.0$ (solid line), $J/t=0.5$ (dashed line), $J/t=1.0$ (dotted line).
}
\label{f-l-w0}
\end{figure}

The role of the magnetic interaction in driving the system towards
an effective antiadiabatic limit is even more evident
when we draw the lattice polaron phase diagram in the $\lambda$-$\omega_0/t$
space (Fig.~\ref{f-l-w0}). For zero exchange coupling $J/t=0.0$
(solid line) it is possible to
distinguish an adiabatic regime, where the lattice polaron
formation is ruled by the condition
$\lambda \bsim 1$, and an antiadiabatic regime
where the polaron occurs for $\alpha^2 \ge 1$
By switching on the magnetic interaction
the polaron crossover approaches the line
$\alpha^2 = 1$ and the validity of the antiadiabatic criterion
is extended for smaller values of $\omega_0/t$.

We can now address
the open issue concerning the modification of the
lattice polaron criterion in the presence of electronic correlation.
The point is to determine whether the simple relation
$\lambda < \lambda_c$ in the adiabatic regime
could be generalized by introducing
properly scaled parameters.
Two alternative pictures have been debated in literature.
According the first one\cite{capone} the relevant parameter in the presence
of electron-electron and magnetic interaction is the ratio
between the lattice polaron energy $g^2/\omega_0$
and the purely electronic ground state energy in the absence
of hole-phonon interaction $E_{0,{\rm mg}} \equiv E_0(\lambda=0)$
\begin{equation}
\lambda^*_1 =\frac{g^2}{\omega_0 |E_{0,{\rm mg}}|}.
\label{crit1}
\end{equation}
An alternative point of view\cite{wellein} regards the effective
hopping amplitude $t^*_{\rm mg}=t^*(\lambda=0)$ as the main
renormalization effect of the exchange coupling
\begin{equation}
\lambda^*_2 =\frac{g^2}{\omega_0 |t^*_{\rm mg}|}.
\label{crit2}
\end{equation}
According these two ideas
the relation $\lambda = \lambda_c$ should be replaced in the presence of
magnetic interaction by $\lambda^* = \lambda_c$.

We have carefully checked the validity of these two
criteria within our exact solution in infinite dimension. We found
that both of them fail to locate correctly the polaron crossover in the
presence of
magnetic coupling because the effective adiabatic ratio is
increased by the decrease of kinetic energy ($t^*_{\rm mg}$)
due to magnetic localization.
This drives the system toward an anti-adiabatic regime in which the polaron
crossover is ruled by the multiphonon constant $\alpha$ which is not
renormalized by magnetic coupling. A naive way to take into account the
reduction of the kinetic energy is to renormalize also the adiabatic
ratio $\omega_0/t$ in similar way with Eqs.~(\ref{crit1})-(\ref{crit2}),
respectively
$\omega_0/|E_{0,{\rm mg}}|$ and $\omega_0/|t^*_{\rm mg}|$.
We have checked also this criteria and we found
that the lattice polaron crossover in the presence
of magnetic interaction can not be
described in a satisfactory way even within this scheme,
although it provides a better agreement than the simple
renormalization of $\lambda$. This just means that the magnetic and lattice
degrees of freedom can not be separated, namely that the
single hole Holstein-$t$-$J$
here considered can not be mapped in a simple Holstein model with
renormalized parameters.

After having analyzed the process of lattice polaron formation,
we can now investigate the properties of the spin polaron
in the presence of hole-phonon interaction.
We have just seen that
the hole-phonon and magnetic interactions are not in competition. On the contrary
the exchange coupling results to favor the lattice polaron formation.
On the same foot we can expect that similar arguments hold
for magnetic polaron formation, namely that
the electron-phonon trapping favors small spin polarons.

\begin{figure}
\centerline{\psfig{figure=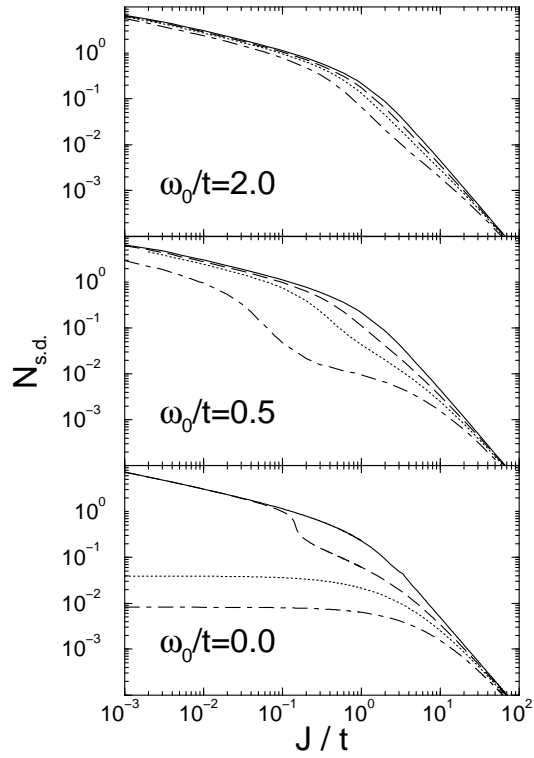,width=7cm,clip=}}
\narrowtext
\caption{Mean number of spin defects
$N_{\rm s.d.}$ as function of the exchange coupling $J/t$ for different
adiabatic parameters: $\omega_0/t=0.0, 0.5, 2.0$ and
$\lambda=0$ (solid lines), $\lambda=0.5$ (dashed lines),
$\lambda=1.0$ (dotted lines), $\lambda=2.0$ (dot-dashed lines).}
\label{f-nsd}
\end{figure}

In Fig.~\ref{f-nsd} we show the mean number of spin defects
$N_{\rm s.d.}$ as function of the exchange coupling $J/t$ for various
electron-phonon couplings $\lambda$ and different adiabatic ratios
$\omega_0/t$. In the pure $t$-$J$ model (solid line) the small-large
magnetic polaron crossover is denoted by a change of slope at about $J/t=1$
which separates a $N_{\rm s.d.} \propto J^{-1}$
from a $N_{\rm s.d.} \propto J^{-1/3}$ regime.\cite{barnes}

This trend is qualitatively unaffected in the anti-adiabatic
regime $\omega_0/t=2.0$ for the electron-phonon coupling here
considered $\lambda \leq 2$. In this situation the lattice polaron
formation occurs as a smooth crossover with negligible
localization. Large-small spin polaron formation is driven therefore
only by the magnetic interaction without any significant interplay
between lattice and spin degrees of freedom. We can schematize
this scenario as a two phases transition,
large magnetic polaron/small magnetic polaron.

The scenario changes approaching the adiabatic limit.
In the intermediate regime $\omega_0/t=0.5$ we can distinguish
a weak electron-phonon coupling
regime ($\lambda \leq 1$) where the magnetic interaction is still
the only relevant energy scale for the large-small spin polaron transition,
and the strong coupling regime ($\lambda > 1$) where lattice polaron
formation interferes with the magnetic one.
The case $\lambda = 2$ (dot-dashed line) is representative of this regime.
For very small $J/t \rightarrow 0$ we recover the usual
$N_{\rm s.d.} \simeq c\, J^{-1/3}$ behavior, characteristic of the large
magnetic polaron. Effective electron-phonon coupling is not sufficient to give
the lattice polaron localization and it solely gives
a reduction of the prefactor $c$ due to the renormalization
of the local hopping amplitude.
For larger $J$, $0.05 \lsim J/t \lsim 5$, we find a remarkable
decrease of the mean number of spin defects $N_{\rm s.d.}$. The origin
of such a decrease is not magnetic since in this region
$N_{\rm s.d.}$ depends only weakly on the exchange coupling $J$.
We can identify this regime as a small magnetic polaron {\em induced}
by lattice polaron trapping.
Finally, for larger interaction $J/t > 5$ the magnetic energy becomes
strong enough to overcome the lattice polaron localization and
we recover a pure magnetic trapping.
The overall evolution from large magnetic polaron to small
magnetic polaron can be described as two crossovers:
large magnetic polaron/small (magnetic) polaron induced by lattice
polaron localization/small magnetic polaron.

Crossovers become even more marked as approaching the adiabatic regime
and become a {\it discontinuous} transition in the
adiabatic limit ($\omega_0/t=0.0$ in Fig.~\ref{f-nsd}).
Detailed calculations for this particular limit
(static lattice distortions) have been explicitly carried
out in Appendix \ref{appadiab}.
The dependence of $N_{\rm s.d.}$ on $J/t$ is drastically different
in the weak $\lambda < \lambda_c$ and in the strong
$\lambda > \lambda_c$ coupling cases. In this latter case
in particular the particles are almost perfectly trapped
the intermediate region of small magnetic polaron induced by
lattice trapping extends towards $J/t=0$.
In the adiabatic limit, for $\lambda>0.844$
we have always small lattice/magnetic polaron (see Appendix \ref{appadiab}).

In the above discussion particular care needs to be paid in
distinguishing between the character (small/large)
and the nature (magnetic of phononic) of
polaron transition.
We can find a similarity between the small/large magnetic polaron
and the small/large lattice polaron transition: both of them
depend on the local probability of the hole to hop from
a site $i$ to an other site $j$. A suitable quantity to express this concept
is the effective hopping amplitude
$t^*$.\cite{ciuchi,wellein,feinberg}
Small values of $t^*/t$ denote strong localization of the hole,
{\it regardless its specific origin} (lattice polaron trapping or
small magnetic polaron).

\begin{figure}
\centerline{\psfig{figure=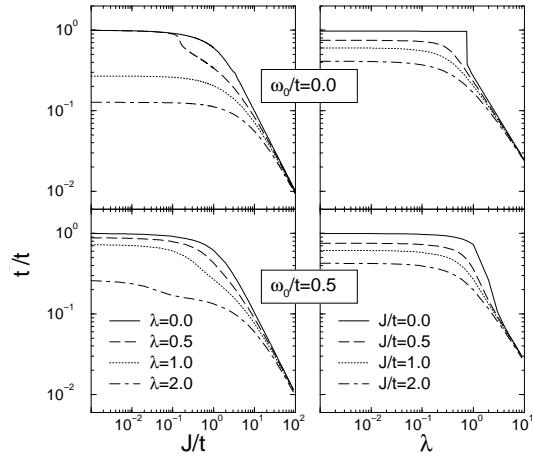,width=7cm,clip=}}
\narrowtext
\caption{Effective hopping amplitude $t^*/t$
as function of the exchange coupling and of
the electron-phonon coupling $\lambda$.}
\label{f-ek-tot-w0=05}
\end{figure}

In Fig.~\ref{f-ek-tot-w0=05} we plot $t^*/t$ as function
of the exchange coupling $J/t$ (left panel)
and of the electron-phonon coupling $\lambda$ (right panel)
for the cases $\omega_0/t=0.0$ and $\omega_0/t=0.5$.
Comparing the left panel of Fig.~\ref{f-ek-tot-w0=05}
with the corresponding cases in Fig.~\ref{f-nsd}
we can clearly identify the trends discussed above.
For $\lambda \lsim 1$ the electron-phonon interaction induces
only a weak reduction of $t^*/t$ while
the localization transition is essentially driven by the
magnetic coupling $J/t$. When the electron-phonon coupling is strong enough
($\lambda \bsim 1$) however the hole dynamics is strongly
suppressed already at $J/t=0$ by lattice polaron trapping,
and higher value of $J/t$ are needed to further decrease the effective
hopping amplitude by magnetic effects.

This behavior is quite similar when we plot $t^*/t$ as function of
$\lambda$ for different exchange coupling $J/t$ (left panel).
We note however that the decrease of the kinetic energy
at $\omega_0/t=0.5$
is steeper when induced by lattice polaron formation than by the magnetic one.
This difference is amplified by approaching the adiabatic regime
$\omega_0/t \ll 1$. In the adiabatic limit  $\omega_0/t=0$
a discontinuous transition occurs at small value of $J/t$ around
a critical value of the coupling which in the limit $J/t=0$ approaches the value
found in the Holstein model\cite{ciuchi} (see Appendix \ref{appadiab}).

\begin{figure}
\centerline{\psfig{figure=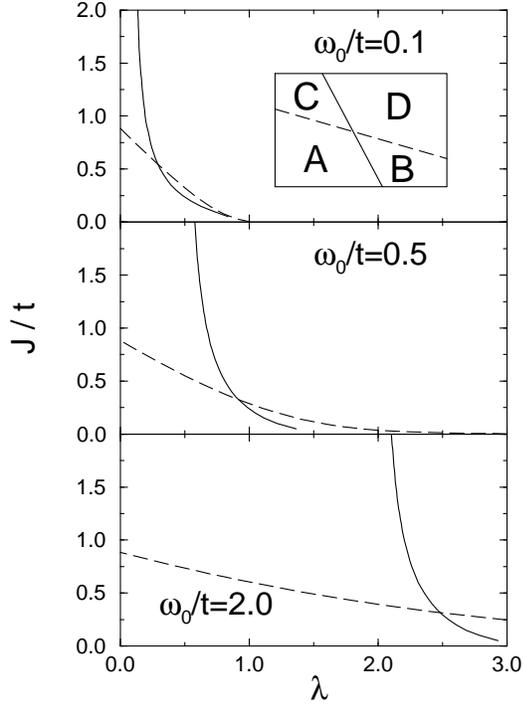,width=7cm,clip=}}
\narrowtext
\caption{Polaron phase diagram of the Holstein-$t$-$J$ model
for different values of the adiabatic ratio $\omega_0/t$.
Solid lines mark the lattice polaron formation, dashed lines the large/small
spin polaron transition. Inset: pictorial sketch of the generic
phase diagram.}
\label{phs}
\end{figure}

We can now summarize the above study in a global polaronic phase
diagram for the Holstein-$t$-$J$ model, shown in
Fig.~\ref{phs}. The solid line marks the lattice polaron
formation according the criterion (\ref{lpcrit}) and the dashed line
the small/large spin polaron transition  by $N_{\rm s.d.}=0.5$.
The dependence of the lattice polaron formation on the magnetic exchange
$J/t$ (solid line) and of the spin polaron transition on electron-phonon
coupling $\lambda$ (dashed line) points out the strong interplay between
the two kind of processes. In particular they are not competing
but sustaining each other.
We can distinguish four regions characterized as follows (see
inset in Fig.~\ref{phs}):
(A) no lattice polaron, large spin polaron; (B) lattice polaron,
large spin polaron; (C) no lattice polaron, small spin polaron;
(D) lattice polaron, small spin polaron.

It is interesting to compare the evolution of the phase diagram
with respect to the
adiabatic ratio $\omega_0/t$. For large $\omega_0/t$ the lattice polaron
formation is not accompanied by a strong hole trapping but
appear as a smooth crossover just like the magnetic transition.
We can thus identify finite regions (B) and (C) where lattice and spin
polaron can be established independently of each other.
Approaching the adiabatic regime ($\omega_0/t=0.1$) the phases
(B) and (C) gradually shrink.

\begin{figure}
\centerline{\psfig{figure=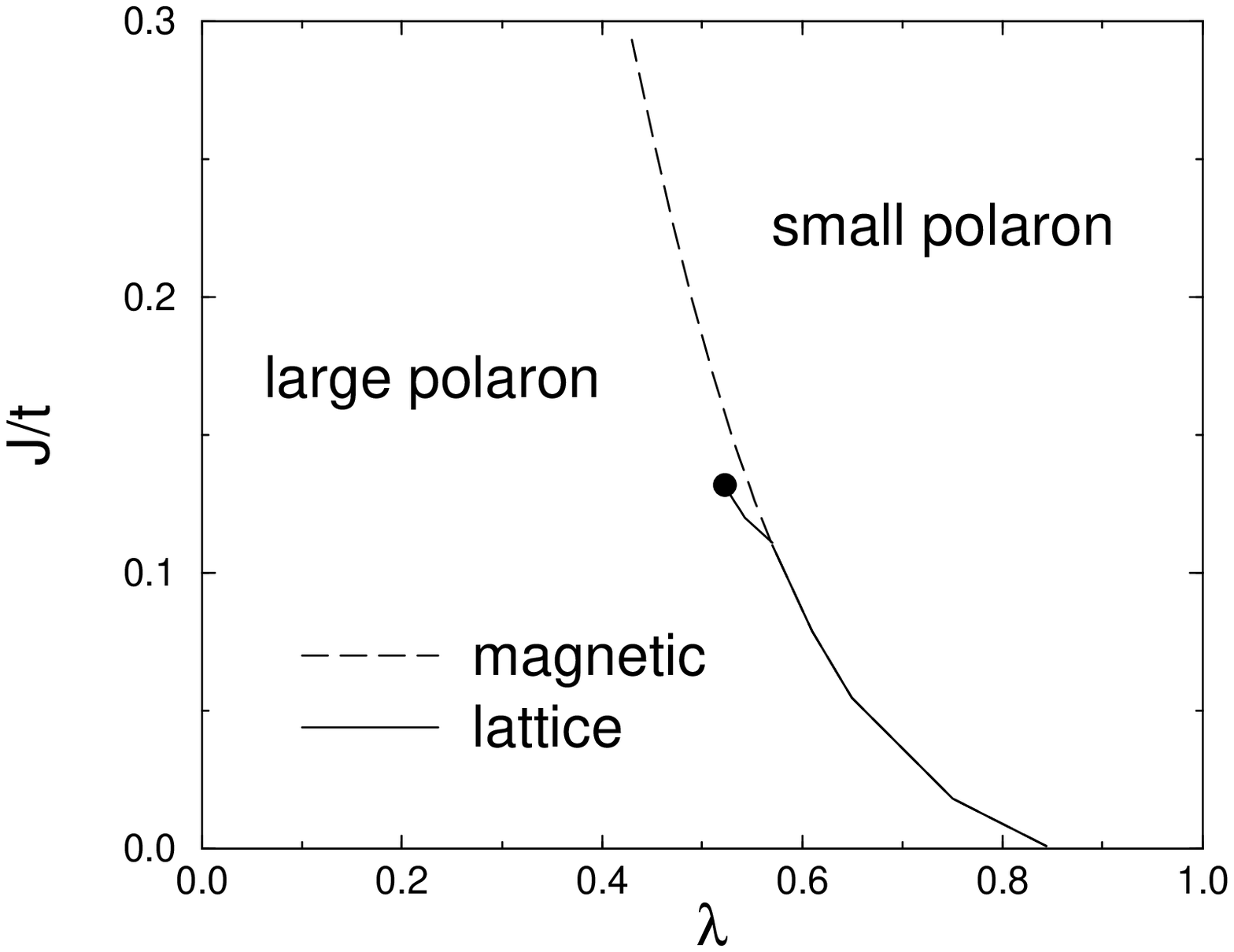,width=7cm,clip=}}
\narrowtext
\caption{Adiabatic phase diagram obtained at $\omega_0/t=0$. 
The solid line marks the large/small lattice polaron discontinuous transition,
the dashed line the large/small spin polaron transition.
The sharp large/small lattice polaron transition disappears
at $J_c/t \simeq 0.132$ (marked by the filled circle)
where it becomes a continuous crossover.}
\label{jc-ad}
\end{figure}

Particular care is needed in the strict adiabatic limit $\omega_0/t=0$.
The criterion described Eq. (\ref{lpcrit}) states the existence of a
multiphonon state as a small polaron key feature.
Due to the localized nature of the system,
lattice distortions with vanishing quantum fluctuation are
always present for any finite $\lambda$ (see Appendix \ref{appadiab}).
This classical lattice state is indeed constituted by an infinite
number of phonon giving $\lambda_c=0$ by using the criterion of
Eq. (\ref{lpcrit}).
Nevertheless we can always identify a discontinuous transition
from very small to large lattice distortions
which survives up to $J/t \simeq 0.132$.
Explicit results are shown in Fig. \ref{jc-ad}. It is thus this transition
which strongly reduces the electron local hopping $t^*$ and enforces the
spin polaron. For larger $J/t$ such a sharp transition disappears
and also the small magnetic polaron formation becomes a smooth crossover
(dashed line in Fig. \ref{jc-ad}).

It should be stressed however that the extension of the strict
adiabatic to realistic finite adiabatic ratio is
relevant as far as quantum fluctuations
are small with respect to the average lattice distortion.
In the case of $\omega_0/t=0.1$ for instance
quantum fluctuations are larger than the small average lattice
distortion found in the adiabatic limit. In this case the multiphonon
criterion as described by Eq. (\ref{lpcrit}) marks as well
the small polaron crossover.

\section{Conclusions}

We have mapped the half-filled Holstein-$t$-$J$ model on an antiferromagnetic
background into a one-particle Hamiltonian (hole interacting with phonons and
spin defects) which can be exactly solved in infinite dimension in terms of a
continued fraction. The method immediately gives access to the hole
spectral density and ground state properties. The main results of our work can
be summarized as:

i) Magnetic and phononic excitations are well separated in the anti-adiabatic
regime and/or in the strong magnetic regime. In this case
phononic peaks are separated by the {\it} bare phonon frequency $\omega_0$.

ii) Magnetic and lattice correlations sustain each other.
Not only polaron crossover is shifted to lower coupling
by magnetic correlations as noticed in
Refs.~\onlinecite{wellein,capone}, but also spin defects are reduced by polaronic
effects.

iii) Phonon retardation is affected for large $J/t$ by magnetic correlations
leading the system
towards anti-adiabatic conditions in which the relevant electron-phonon
coupling changes
from $\lambda$ to $\alpha$. Therefore it is not sufficient to scale $\lambda$
with the renormalized electron kinetic energy to locate the polaron crossover.

iv) We identify a crossover region between regions of parameters space
in which magnetic and
lattice polaron occurs independently (B-C) and regions (A-D) in which they are
mutually dependent. We find the rigorous adiabatic regime to be quite peculiar
wherein the small magnetic polaron
formation is always accompanied by a small polaron formation.

The main drawback of our DMFT method lies in neglecting dispersion of spin waves
which leads to hole coherent motion. This is evident in our spectra which are
constituted by $\bf k$-independent peaks. As a starting point
to overcome this difficulty we explicitly
included in Eq. (\ref{hamilha}) terms which would lead to hole
delocalization as well as to spin wave dispersion in the next order in $1/z$.
A controlled way to include this processes
is currently under investigation. We expect that taking into account the
coherent quasiparticle motion of the hole due to the quantum
spin fluctuation would modify the low energy features of
the spectral function by giving rise to finite bands.
However we think that the gross features of the spectral weight
would not be strongly affected. In addition
the existence of spin fluctuations could lead to delocalization
of the large polaron found in the adiabatic limit
at weak coupling.

We acknowledge useful discussion with M. Capone.

The authors acknowledge support
of the Italian Ministry of University and Scientific Research fund cofin-99.

\appendix

\section{Exact solution of the one hole Holstein-\lowercase{$t$}-$J$ model
in infinite dimension}
\label{appfrac}

Let us consider the Hamiltonian in Eq. (\ref{hamilhatjz}) which
we write as $H = H_t + H_L$. Here $H_t$ contains the hopping terms
and $H_L$ all the other {\em local} contribution.
In the absence of any hole, the ground state is just the antiferromagnetic
one which can be written as
$\left| AF \right\rangle
=
\left| 0;0 \right\rangle$,
where $\left| 0;0 \right\rangle$ represents the antiferromagnetic
background with no phonon and no spin defect on any site.
In similar way we can introduce the notation
\begin{eqnarray}
&&\left| n_{(i)}, m_{(j)}, \ldots ; l, k, \ldots \right\rangle
\nonumber\\
&&\hspace{1cm}
\equiv
\left[
\frac{(b_i^\dagger)^n}{\sqrt{n}}
\frac{(b_j^\dagger)^m}{\sqrt{m}}
\ldots
\right]
\left[
a_l^\dagger
a_k^\dagger
\ldots
\right]
\left| 0;0 \right\rangle
\label{diff}
\end{eqnarray}
to express the state with $n$ phonon on the site $i$, $m$ phonons
on the site $j$, and defects of spin on the site $l$, $k$, \ldots

Aim of investigation of this appendix will be the Green's function
\cite{notelocal}
\begin{equation}
G_{ii}(\omega)=
\left\langle 0;0 \left|
h_i \frac{1}{\omega-H} h_i^\dagger
\right| 0;0 \right\rangle,
\label{greenl}
\end{equation}
which can be considered as the $(0,0)$ element [$G(\omega) = G^{00}(\omega)$]
of the generalized
Green's function
\begin{equation}
G^{nm}_{ii}(\omega)=
\left\langle 0 ;  n_i \left|
h_i \frac{1}{\omega-H} h_i^\dagger
\right| m_i ; 0 \right\rangle.
\label{green-nm}
\end{equation}

In addition we introduce the {\em atomic}
propagator
\begin{equation}
g^{nm}_{ii}(\omega)=
\left\langle 0 ;  n_i \left|
h_i \frac{1}{\omega-H_L} h_i^\dagger
\right| m_i ; 0 \right\rangle,
\label{greenat}
\end{equation}
which satisfies the following properties ($i \neq j$):
\begin{eqnarray}
&&\left\langle 0 ;  n_i,m_j \left|
h_j \frac{1}{\omega-H_L} h_j^\dagger
\right| p_j,n_i ; 0 \right\rangle
\nonumber\\
&&\hspace{1cm}=
\left\langle 0 ;  m_j \left|
h_j \frac{1}{\omega-H_L-n\omega_0} h_j^\dagger
\right| p_j ; 0 \right\rangle
\nonumber\\
&&\hspace{1cm}=
g^{mp}_{jj}(\omega-n\omega_0),
\label{pro2}
\end{eqnarray}
and
\begin{eqnarray}
&&\left\langle s_i ;  p_i \left|
h_j
\frac{1}{\omega-H_L}
h_j^\dagger
\right| q_i ; s_i \right\rangle
\nonumber\\
&&\hspace{1cm}=
\left\langle 0 ;  p_i \left|
h_j
\frac{1}{\omega-J/2-H_L}
h_j^\dagger
\right| q_i ; 0\right\rangle
 \delta_{p,q}
\nonumber\\
&&\hspace{1cm}=
\left\langle 0 ;  0 \left|
h_j
\frac{1}{\omega-J/2-H_L-p\omega_0}
h_j^\dagger
\right| 0 ; 0\right\rangle
 \delta_{p,q}
\nonumber\\
&&\hspace{1cm}=
g^{00}_{jj}(\omega-J/2-p\omega_0)\delta_{p,q}.
\label{pro3}
\end{eqnarray}
Eqs. (\ref{pro2},\ref{pro3}) stem from the fact that the
electron-phonon coupling in $H_L$ is operative only on the site on which
the hole stays.

Let us now expand in Eq. (\ref{greenl})
the resolvent $1/(\omega-H)$ in powers of $H_t$:
\begin{eqnarray}
\frac{1}{\omega-H}
&=&
\frac{1}{\omega-H_L}
+\frac{1}{\omega-H_L}H_t\frac{1}{\omega-H_L}
\nonumber\\
&&+\frac{1}{\omega-H_L}H_t\frac{1}{\omega-H_L}H_t\frac{1}{\omega-H_L}
+\ldots
\label{expans}
\end{eqnarray}

It is easy to see that all the odd powers do not contribute in Eq. (\ref{greenl})
since they create or destroy a odd number of spin defects.
More generally, since the initial and final states do not contain
spin defects, it is clear that all the spin defects created in the dynamics
must be destroyed on each site before to reach the final state.
In infinite dimension this constraint selects only the retraceable paths.
Each ``step forward'' is thus ruled by the term
$t/(2\sqrt{z}) \sum_{\langle \alpha \beta \rangle}
h_\alpha^\dagger a_\alpha h_\beta$ and each ``step backward''
by its complex conjugate.

At the second order in $t$ we have for instance:
\begin{eqnarray}
&&\left[G^{(2)}_{ii}(\omega)\right]^{nm}
\nonumber\\
&&\hspace{0.3cm}=\left\langle 0 ;  n_i \left|
h_i
\frac{1}{\omega-H_L}H_t\frac{1}{\omega-H_L}H_t\frac{1}{\omega-H_L}
h_i^\dagger
\right| m_i ; 0 \right\rangle
\nonumber\\
&&
\hspace{0.3cm}=
\sum_{\langle \alpha,\beta \rangle}
\sum_{\langle \gamma,\delta \rangle}
\left\langle 0 ;  n_i \left|
h_i
\frac{1}{\omega-H_L}
\frac{t}{2\sqrt{z}}h_\alpha^\dagger a_\alpha h_\beta
\frac{1}{\omega-H_L}
\right.\right.
\nonumber\\
&&\left.\left.
\hspace{2cm}\times
\frac{t}{2\sqrt{z}}h_\gamma^\dagger a_\delta^\dagger h_\delta
\frac{1}{\omega-H_L}
h_i^\dagger
\right| m_i ; 0 \right\rangle.
\label{2or}
\end{eqnarray}

Eq. (\ref{2or}) can be expressed in terms of the atomic propagator $g$
by introducing the identity operator in each hopping term:
$h_\alpha^\dagger a_\alpha h_\beta
= \sum_M h_\alpha^\dagger a_\alpha |M\rangle \langle M|
h_\beta$, $h_\gamma^\dagger a_\delta^\dagger h_\delta
= \sum_N h_\gamma^\dagger |N\rangle \langle N|
a_\delta^\dagger h_\delta$ where $|M\rangle$, $|N\rangle$
are complete sets of states.
It is easy to check that only the states
$\sum_M |M\rangle \langle M|=
\sum_{p} |p_i;s_i\rangle \langle s_i,p_i|$
and $\sum_N |N\rangle \langle N|
= \sum_{q} |q_i;s_i\rangle \langle s_i,q_i|$ give a non-zero contribution.
Using the properties Eqs. (\ref{pro2})-(\ref{pro3}) we have thus:
\begin{equation}
\left[G^{(2)}_{ii}(\omega)\right]^{nm}
=-g^{np}_{ii}(\omega)
\frac{t}{2}g^{00}_{jj}(\omega-J/2-p\omega_0)\frac{t}{2}g^{pm}_{ii}(\omega).
\end{equation}

Applying a similar procedure for all the order of the expansion
(\ref{expans}) we obtain a perturbative expression for
the Green's function $G$:
\begin{eqnarray}
G^{nm}_{ii}(\omega)&=&
g^{nm}_{ii}(\omega)-
g^{np}_{ii}(\omega)\frac{t}{2}g^{00}_{jj}(\omega-J/2-p\omega_0)
\frac{t}{2}g^{pm}_{ii}(\omega)
\nonumber\\
&&+
g^{np}_{ii}(\omega)\frac{t}{2}g^{00}_{jj}(\omega-J/2-p\omega_0)\frac{t}{2}
\left[g^{pq}_{ii}(\omega)
\frac{t}{2}g^{00}_{kk}(\omega-J/2-q\omega_0)\frac{t}{2}g^{qm}_{ii}(\omega)\right]
\nonumber\\
&&+
g^{np}_{ii}(\omega)\frac{t}{2}\left[g^{0q}_{jj}(\omega-J/2-p\omega_0)\frac{t}{2}
g^{00}_{kk}(\omega-2 J/2-p\omega_0-q\omega_0)\frac{t}{2}
g^{q0}_{jj}(\omega-J/2-p\omega_0)\right]
\frac{t}{2}g^{pm}_{ii}(\omega)+\ldots,
\label{expans2}
\end{eqnarray}
which can be resummed in the compact form (we drop now the site indices):
\begin{eqnarray}
G^{nm}(\omega)&=&
g^{nm}(\omega)
-g^{np}(\omega)\frac{t}{2}G^{00}(\omega-J/2-p\omega_0)
\frac{t}{2}G^{pm}(\omega),
\label{fin2}
\end{eqnarray}
or in the matricial form:
\begin{equation}
\left[
{\bf G}^{-1}(\omega)\right]^{nm}
= \left[{\bf g}^{-1}(\omega)\right]^{nm} -\delta_{n,m}
\frac{t^2}{4}G^{00}(\omega-n\omega_0-J/2).
\label{tt}
\end{equation}

In particular the terms in square brackets in Eq. (\ref{expans2})
correspond to the iteration up to the second order respectively of $G^{pm}(\omega)$
and of
$G^{00}(\omega-J/2-p\omega_0)$ according to Eq. (\ref{fin2}).

Note that the contribution of the hopping processes
[second term in Eq. (\ref{tt})] is diagonal in the phonon space.
This is due to the property (\ref{pro3}) which relies on the locality
of the electron-phonon interaction.

From the explicit solution of the atomic problem:
\begin{equation}
\left[{\bf g}^{-1}(\omega)\right]^{nm}
= (\omega-n\omega_0)\delta_{n,m} +gX^{nm},
\end{equation}
where
\begin{equation}
X^{nm}=\sqrt{m+1}\delta_{n,m+1}+\sqrt{m}\delta_{n,m-1},
\end{equation}
we end up finally with following self-consistent equation
for the Green's function in the multiphonon space:
\begin{equation}
\left[{\bf G}^{-1}(\omega)\right]^{nm}
= gX^{nm} + \delta_{n,m}
\left[\omega-n\omega_0 - \frac{t^2}{4}G^{00}(\omega-n\omega_0-J/2)\right].
\label{tri}
\end{equation}

The solution of Eq. (\ref{tri}) reduces to the inversion problem of
a tridiagonal matrix\cite{Viswanath-Muller,ciuchi}.
The diagonal elements $G^{nn}$ can be expressed as continued
fractions\cite{ciuchi} obtaining Eqs. (\ref{eqG})-(\ref{gt})
and (\ref{Gnn})-(\ref{Sigma_abs}).
Eq. (\ref{tri}) looks similar to Eq. (34) of Ref. \onlinecite{ciuchi}
with the important difference that $G^{-1}_0$ depends now
on the exchange energy $J$.
The same result can be obtained for $G^{00}$ using
diagrammatic techniques as in Ref. \onlinecite{strack}.


\section{Adiabatic limit}
\label{appadiab}

In this Appendix we will solve the problem of one electron moving in an
infinite coordination {\it static} lattice.
Here we follow the derivation of the adiabatic limit done in the Holstein model
in Ref.~\onlinecite{ciuchi} and we use the same notations.
The adiabatic limit is achieved as $M\rightarrow\infty$ keeping
$k=M\omega^2_0$ constant.
The coupling constant of Hamiltonian Eq. (\ref{hamilhatjz})
is given in terms of $g^\prime$ by $g=g^\prime/\sqrt{2M\omega_0}$.
The polaron energy $\epsilon_p = -g^2/\omega_0 = -g^{'2}/2k$
is then a well defined quantity in the adiabatic limit.
Minimizing the ground state energy of
the Holstein model with respect to the lattice deformation $X_i$ 
around a given site $i$ we have
\begin{equation}
\label{X-n-adiab}
X_i = \frac{g^\prime}{M\omega^2_0} \langle n_i \rangle.
\end{equation}
Therefore charge localization around a given
site means also a localization of lattice deformations.

The infinite coordination limit together with Eq. (\ref{X-n-adiab}) implies
that, for a single hole, only one site is appreciably distorted. Around a
localization center (site $0$) the nearest neighbor deformation
is of the order $O(1/z)$,
the next nearest neighbor deformation ($O(1/z^2)$) and so on, so that the total
charge can be spread over several shells of neighbors even in the
$z\rightarrow\infty$ limit but the deformations around a localization center
vanishes in the $z\rightarrow\infty$ limit.

From the equation of motion we can derive a hole propagator for a {\it given}
set of lattice deformations.\cite{ciuchi}
The main simplification of the
$d\rightarrow\infty$ limit
is then that the elastic energy {\it is solely determined by the
$0$-site deformation}, for it depends on $X^2_i$. Consequently we
have two kinds of local propagators:
one which describes the
motion of the electron from site $0$ back to site $0$
and which depends upon the deformation:
\begin{equation}
\label{Goo-ad}
G_{00}(\omega) =  \frac{1}{\omega+g^\prime X_0-\Sigma_{\rm hop}(\omega)};
\end{equation}
and a second propagator which enters in $\Sigma_{\rm hop}(\omega)$
[Eq. (\ref{Sigmat})] which
does not depend on lattice
deformation but {\it depends on exchange $J$}:
\begin{equation}
\label{G11-ad}
G(\omega) = \frac{1}{\omega -
\frac{\displaystyle t^2}{\displaystyle 4}G(\omega-
J/2)}.
\label{green}
\end{equation}
It is worth to note that Eqs. (\ref{Goo-ad},\ref{green}) can be obtained within
$t-J-v_i$ model of ref. \cite{strack} with on-site energy $v_1=g^\prime X_0$
and zero neighbors energy.

The lowest energy pole of $G_{0,0}$ gives the electronic energy $E_{\rm el}$
for a given $0$-site deformations $X_0$:
\beq
E_{\rm el}+g^\prime X_0-\frac{t^2}{4}\mbox{Re}\left[
G\left(E_{\rm el}-\frac{\displaystyle J}{\displaystyle 2}\right)\right]=0.
\label{pole-ad}
\eeq 

Eq. (\ref{pole-ad}) also defines $X_0$ as a function of $E_{\rm el}$. By defining
properly scaled deformation
$X_0=g^\prime u/k$ and energies $E_{\rm el}=t \epsilon$, $E_{\rm tot}=t \epsilon_{\rm tot}$ 
and exploiting the continued fraction structure\cite{strack} of Eq. (\ref{G11-ad}) 
we have
\beq
\label{deformation-ad}
u(\epsilon) = \frac{1}{2\lambda} \left (  \frac{1/4}{
\displaystyle \epsilon-J/2-\frac{\displaystyle 1/4}{
		\displaystyle \epsilon -J-\frac{\displaystyle 1/4}{
			\displaystyle \epsilon -3J/2-\ldots 
		}
	}
} - \epsilon \right).
\eeq
By adding the elastic contribution we have the total energy which has
to be minimized
with respect $\epsilon$:
\beq
\label{toten-ad}
\epsilon_{\rm tot}(\epsilon)=\lambda u^2(\epsilon)-\epsilon.
\eeq

The total energy minimization can be carried out explicitly in the strong
coupling limit i.e. when $J/t \gg 1$ or when $\lambda \gg 1$.
In these limits the
continued fraction appearing in Eq. (\ref{deformation-ad}) can be neglected
giving a linear dependence for $u$ ($u=-\epsilon/2\lambda$). The minimization of 
Eq. (\ref{toten-ad}) gives $\epsilon=-2\lambda$ and $\epsilon_{\rm tot}=-\lambda$.
This limit corresponds to a small lattice/magnetic polaron regime. In this case
the deformation ``saturates'' the charge deformation relation of 
Eq. (\ref{X-n-adiab}) and the hole is perfectly localized on a given site.

Another interesting case is the $J/t\rightarrow 0$ limit. An analytical
calculation can be done in this limit following the lines of
Ref.~\onlinecite{ciuchi}.
We have in this case 
a solution with vanishing deformation which gives the lowest energy for
$\lambda < 0.844$. In this case the Green function is the same of
Ref.~\onlinecite{strack} and consists of a semicircular band of
{\it localized} states.
And a 
a solution with non-zero deformation which gives the lowest energy for
$\lambda > 0.844$. In this case a pole emerges out of the band 
at low energies.

The transition at $\lambda_c=0.844$ is found to be discontinuous.
It is important to notice that even if it is possible to follow the formal
steps of Ref.~\onlinecite{ciuchi} to recover these solutions the physical
interpretation of case with vanishing deformation is quite different. 
In particular we may understand
this solution as describing a {\it localized} large lattice/magnetic polaron in
the limit of infinitely large polaronic radius in contrast to the case of the
pure Holstein model where in this case the motion of the electron
is {\it coherent} through the lattice.\cite{ciuchi}
Instead solution associated to a non vanishing deformation
has the same character in both models i.e. it describes a
localized small polaron.

In the general case the minimization of Eq. (\ref{toten-ad}) can  be
easily carried out numerically. Derivatives of the ground state energy with
respect to $g^\prime$ and $J$ gives respectively the hole-phonon and the
exchange (mean number of spin defects) contributions to the total energy. The
hole kinetic energy being obtained by subtraction. These derivatives of the
ground state energy are discontinuous at the transition found for 
$J/t=0$ at $\lambda_c=0.844$. The discontinuous large to small lattice/magnetic
polaron transition exists up to $J/t=0.132$. For larger magnetic couplings 
a smooth crossover takes place.

\newpage

\begin{references}
\bibitem{note2}
Strictly speaking, this is also equivalent to the hole kinetic energy.
However we prefer to call it ``hopping amplitude'' $t^*$
to underline its role in the renormalization of the parameters
$J/t$, $g^2/ \omega_0 t$, $\omega_0 / t$.
\bibitem{corrige}
Note that in Ref. \onlinecite{ciuchi} Eqs. (40)-(42)
must be corrected by a misprint:
$G_0^{-1}(\omega)$ must be replaced everywhere
by $G_0^{-1}(\omega-n\omega_0)$.
\bibitem{notelocal} Although unnecessary in infinite dimension,
we retain for the moment the site indices in order to make
clear the consequences of the retraceable path constraint imposed
by the antiferromagnetic background.
\end{references}
\end{document}